\documentclass[%
 aip,
jmp,%
 amsmath,amssymb,
preprint,%
]{revtex4-1}

\usepackage{graphicx}% Include figure files
\usepackage{dcolumn}% Align table columns on decimal point
\usepackage{bm}% bold math
\usepackage[mathlines]{lineno}% Enable numbering of text and display math
\begin{document}

\preprint{AIP/123-QED}

\title{The effect of time-varying flow-shear on the nonlinear stability of the boundary of magnetized toroidal plasmas}
\author{Youngmin Oh}
\affiliation{Beijing Computational Science Research Center, Beijing 100193, China}
\email{youngminoh19850329@csrc.ac.cn}
\author{Hyung Ju Hwang}
\affiliation{Department of Mathematics, Pohang University of Science and Technology,
Pohang, Gyeongbuk 37673, Republic of Korea}
\email{hjhwang@postech.ac.kr}

\author{Michael Leconte}
\affiliation{National Fusion Research Institute, Daejeon 34133, Republic of Korea
}
\email{mleconte@nfri.re.kr}

\author{Gunsu S. Yun}
\affiliation{Department of Physics, Pohang University of Science and Technology, Pohang,
Gyeongbuk 37673, Republic of Korea}
\email{(corresponding) gunsu@postech.ac.kr}

\date{\today}% It is always \today, today,

\begin{abstract}
We propose a phenomenological yet very general model in a form of generalized complex Ginzburg-Landau equation to understand the dynamics of the quasi-periodic fluid instabilities (called edge-localized modes) in the boundary of toroidal magnetized high-temperature plasmas. The model reproduces key dynamical features of the boundary instabilities observed in the high-confinement state plasmas on the KSTAR tokamak, including quasi-steady states characterized by field-aligned filamentary eigenmodes, transitions between different eigenmodes, and rapid transition to non-modal filamentary structure prior to the relaxation. 
It is found that the inclusion of time-varying perpendicular sheared flow is crucial for reproducing the observed dynamical features. 
\end{abstract}

\pacs{52}% PACS, the Physics and Astronomy
%Use showkeys class option if keyword
\maketitle
\section{\label{sec1}Introduction}
Relaxation phenomena in magnetized plasmas are widespread in nature [\onlinecite{rp1, rp2}]. A notable example is the explosive flares on the surface of the Sun. Another example is the semi-periodic explosive bursts appearing at the boundary of toroidally-confined high-temperature plasmas (e.g., tokamak).
In toroidal magnetic confinement devices, sufficient heating of the plasma can lead to a transition from
low-confinement state ($L$-mode) to high-confinement state ($H$-mode) if the heating power exceeds a threshold. During the
transition, a transport barrier (called pedestal) spontaneously appears at the edge of plasma via
strong $E\times B$ flow shear which reduces heat and particle transports.
However, this barrier is quite unstable and prone to a class of fluid instabilities called edge-localized modes (ELMs) driven by the large gradient of density, temperature, current density, and flow
[\onlinecite{eb3,eb5, eb2,eb4,eb1,eb6}]. 
It is believed that these instabilities are responsible for the relaxation (or crash) of the transport barrier, i.e., rapid expulsion of heat and particles.
The expulsion events are commonly called ELM crash.
The H-mode plasmas are characterized by semi-periodic cycles between slow transport barrier buildup and its fast relaxation.

The ELM crash must be controlled because the natural or uncontrolled crashes induce significant heat and particle fluxes which can damage the plasma-facing walls of the confinement device. 
Magnetic perturbations have been used successfully to mitigate or suppress the crash
[\onlinecite{ex2,ex4,mp3}] but the underlying mechanisms of mitigation and suppression are still unclear.
Accordingly, it is crucial to understand the dynamics of ELM for more reliable and robust methods to avoid the crash. For this reason, a
nonlinear mathematical analysis is required  beyond linear stability analyses [\onlinecite{ls}].

For the purpose of studying the nonlinear behavior, a nonlinear model for the perturbed pressure was derived in a form of complex Ginzburg-Landau equation based on a 1D reduced MHD model [\onlinecite{leconte2016ginzburg}]. The numerical solutions to the model equation showed nonlinear relaxation oscillations with the characteristics of type-III ELM. 
Inspired by [\onlinecite{leconte2016ginzburg}], we mathematically studied the model equation to understand the effect of perpendicular flow shear on the nonlinear behavior of the perturbed pressure during the ELM cycle [\onlinecite{2017arXiv170608036O}].
More precisely, it was shown that there exists a linearly stable symmetric
steady state for small shear and the first eigenvalues of
unstable states for the case of zero shear are bounded below by a positive
constant. 
In the case of large shear, a theoretical clue was found for the long-time behavior of
the solutions: 1) nonlinear oscillation; 2) convergence to $0.$ 
The theoretical results were supported by numerical verifications.

However, in [\onlinecite{2017arXiv170608036O}], the shear strength was set constant in time, which was insufficient to explore clues for the various phenomena observed in experiments on the Korea Superconducting Tokamak Advanced Research (KSTAR) device such as quasi-steady state with a single eigenmode-like structure [\onlinecite{yun2011two}] and fast transitions between the quasi-steady states [\onlinecite{lee2015toroidal}].
In this paper, the effect of time-varying flow shear is analyzed as the key for accessing different dynamical states.
The remaining of the article is organized as follows: in section II, we present the analysis of the model for the case of a single-mode, in section III, we extend the model to treat the case of two coupled modes. In section IV, we discuss the results and give a conclusion.

\section{\label{sec2}Analysis of single-mode}
\begin{figure}
[ptb]
\begin{center}
\includegraphics[
width=\textwidth,
keepaspectratio
]%
{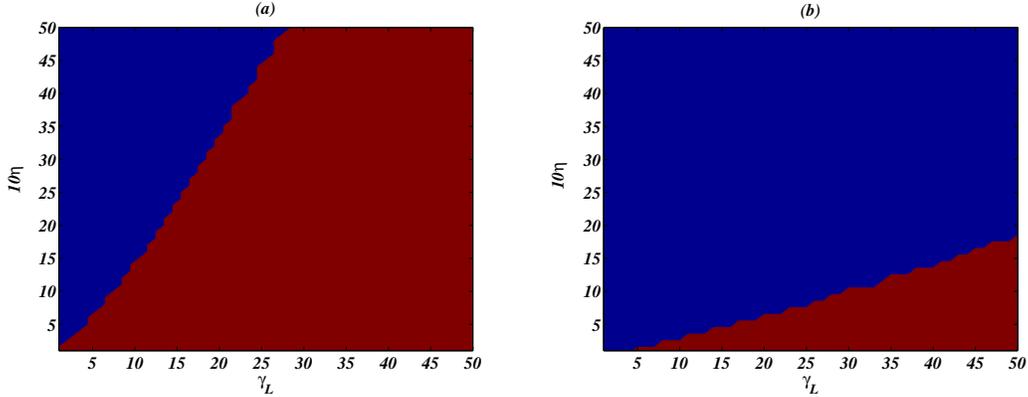}%
\caption{The qualitative long-time behavior of a solution $P\left(  t,x\right)  $ to
Eq.~(\ref{main}): nonlinear oscillation (red regions) or convergence to $0$ (blue
regions) on the Neumann and the Dirichlet boundary conditions in $(a)$ and $(b)$ respectively. Here, we set $\gamma_{N}=1,$ $A=50,$ and $W_{K}\left(  x\right)  =\tanh\left(
25x\right)$. In each case, there exists a clear boundary separating the two regions.}%
\label{det}%
\end{center}
\end{figure}

\begin{figure}
[ptb]
\begin{center}
\includegraphics[
width=\textwidth,
keepaspectratio
]%
{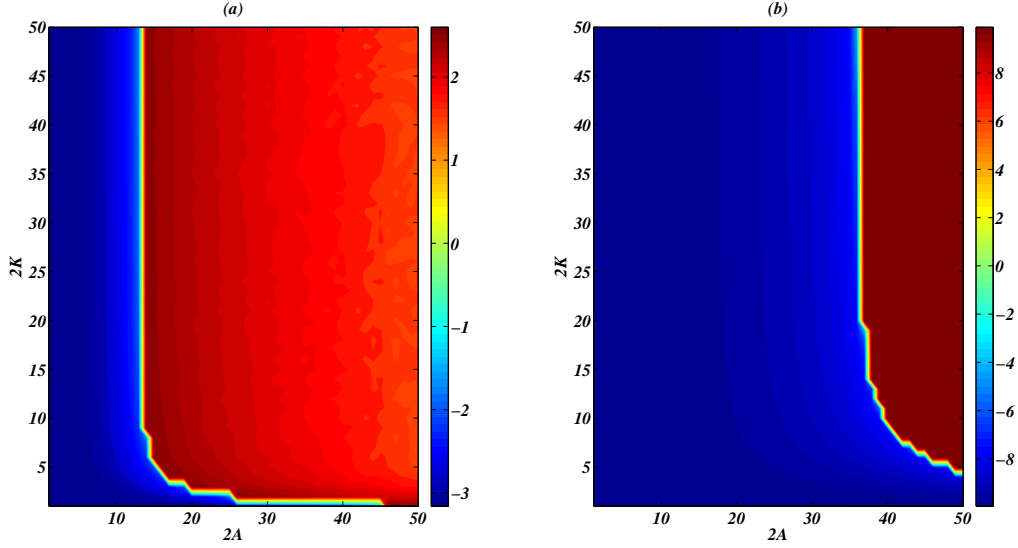}%
\caption{The qualitative long-time behavior of a solution $P(t,x)$ to Eq.~(\ref{main}) with $\eta=1,$ $\gamma_{N}=1,$and $\gamma_{L}=10$: nonlinear
oscillation (red regions) or convergence to a nonzero steady state (blue
regions) for the Neumann and the Dirichlet
boundary conditions in $(a)$-$(b)$ respectively: The values in the red regions in $\left(
a\right)  $-$\left(  b\right)  $ denote $\lim_{t\rightarrow\infty}%
\max\left\vert P\left(  t,0\right)  \right\vert$. The values in the blue regions $\left(  a\right)
$-$\left(  b\right)  $ denote $-\left\vert
P\left(0\right)  \right\vert $ for a nonzero steady state $P(t,x)=P(x)$. It is clear that there is $A_{K}$ for each $K$ which determines the long-time behavior of the solution. Note that, as approaching the interfaces, the values in the red and blue regions increase, so the amplitude of nonlinear oscillations increases and $\vert P(0) \vert$ for a nonzero steady state $P(x)$ decreases, but not to $0$.}%
\label{peak}
\end{center}
\end{figure}

We consider the following single-mode equation for the perturbed pressure $P(t,x,y)$ in cylindrical magnetized plasma assuming local slab geometry with the magnetic field direction $z$, the local radial direction $x$, and the perpendicular direction $y$:%
\begin{equation}
\partial_{t}P+\gamma_{N}\left\vert P\right\vert ^{2}P=iAW_{K}\left(  x\right)
P+\gamma_{L}P+\eta\partial_{x}^{2}P, \label{main}%
\end{equation}
where $W_{K}\left(  x\right)  =\tanh\left(  Kx\right)$
 is the prescribed shear flow with $x\in\left[  -1,1\right]$,
 $K>0$ is the inverse of the shear layer width, and $A \ge 0$ is the shear flow strength.
 Eq.~(\ref{main}) may be considered a generalization of Ginzburg-Landau equation (GLE) with constant complex coefficients.
Note that $P$ represents the complex-valued amplitude of a Fourier mode, i.e. $\delta P(x,y,t) = P(x,t) e^{i k y} +c.c.$.
Here, $\gamma_{N}$, $\gamma_{L}$ and $\eta$ are constant coefficients for the
nonlinear, the linear growth and the dissipative terms respectively. It was
observed that the behavior of a solution to Eq.~(\ref{main}) is completely
different with the presence of the flow-shear for both the Dirichlet and Neumann
boundary conditions [\onlinecite{2017arXiv170608036O,leconte2016ginzburg}]. 
Since it is unclear which boundary condition is reasonable in real experiments, 
both types of boundary conditions are considered here to understand the long-time behavior of a solution $P\left(t,x\right) $ to Eq.~(\ref{main}):%
\begin{align*}
P\left(  t,\pm1\right)   &  =0\text{ (Dirichlet),}\\
\frac{\partial P}{\partial x}\left(  t,\pm1\right)   &  =0\text{ (Neumann)}.
\end{align*}

Inspired by [\onlinecite{2017arXiv170608036O}], we will consider two subjects for the
model Eq.~(\ref{main}). The first subject is to characterize the long-time
behavior of a solution $P(t,x)$ for the fixed large shear strength $A$ 
so that we can distinguish the regions of either convergence to $0$ or nonlinear oscillations 
in the $\gamma_{L}$--$\eta$ parameter space. 
The second subject is to characterize the long-time behavior of a solution $P(t,x)$ between nonlinear
oscillations and convergence to nontrivial steady states in the $A$--$K$ parameter space under suitable fixed parameters $\gamma_{L}$ and $\eta$ such that non-trivial solutions are guaranteed. 
We find a threshold $A_{K}>0$ for each $K$ such
that solutions converge to a nonzero steady state of Eq.~(\ref{main}) for
$A<A_{K}$ and nonlinearly oscillate for $A>A_{K}$. 
Combining these results, we propose that the salient features of the ELM dynamics observed in the KSTAR H-mode plasmas can be explained based on time-varying perpendicular shear flow.

\subsection{Long-time behavior of $P(t,x)$ on $\gamma_{L}$ and
$\eta$}%
Notice that the Dirichlet boundary condition does not allow nonzero uniform
steady states of Eq.~(\ref{main}) even without the shear in contrast with the
Neumann boundary condition. Nevertheless, we obtained similar results for both
boundary conditions. Fig.~\ref{det} represents the long-time behaviors of a
solution $P(t,x)$ on $\gamma_{L}$ and $\eta$ for a fixed large
$A=50$ in both boundary conditions. The blue regions in Fig.~\ref{det}
(a)--(b) display that $P(t,x)$ converges to $0$ as $t\rightarrow\infty$. 
Conversely, red regions in Fig. \ref{det} $(a)$-$(b)$
display that $P\left(  t,x\right)  $ oscillates nonlinearly in time. These
results show a certain relation between $\eta$ and $\gamma_{L}$ which
determines the long-time behavior of $P\left(  t,x\right)  $. 
Inspecting Fig. \ref{det}, it is clear that nonlinear oscillations are guaranteed only
if the ratio $\gamma_{L}/\eta$ is sufficiently large. Otherwise, $P(t,x)$ converges to $0$.
Note that the parameters in Eq.~(\ref{main}) are related to heat flux $Q$ as (see [\onlinecite{leconte2016ginzburg}]), %
\begin{align*}
\gamma_{L} = \gamma_{L0}\frac{Q-Q_{c}}{\eta}a p_0^{-1},\text{ } \gamma_{N}=\frac{a^2\gamma_{L0}^{2}}{\eta}, \text{ and } \frac{\gamma_{L}}{\eta}  &  \propto\gamma_{L0}\frac{Q-Q_{c}}{\eta^{2}}.
\end{align*}
where $Q_c$ is the threshold heat flux related to the critical pressure gradient for linear instability, $p_0$ is the reference pressure, and $a$ denotes the radius of the cylinder (see [\onlinecite{leconte2016ginzburg}] for detail). 
Therefore, even if the heat flux $Q$ exceeds the linear threshold $Q_{c}$, nonlinear
oscillations may not occur if $0 < Q-Q_{c} \ll 1$ such that 
$\gamma_{L} \ll 1$ and $\left(\gamma_{L}/\eta\right) \ll 1.$ 
This is consistent with experiment observations
since it is known that ELM crash does not immediately occur after $Q$ exceeds
$Q_{c}$ (see Fig. 1 in [\onlinecite{schmitz2012role}]). It is also possible to
interpret the case of $Q-Q_{c}<0$ ($\gamma_{L}<0)$ as $L$-mode. $\gamma_{L}<0$
guarantees the long time behavior of $P\left(  t,x\right)  $ such that
$\lim_{t\rightarrow\infty}\left\vert P\left(  t,x\right)  \right\vert
\rightarrow0.$ Therefore, Eq.~(\ref{main}) provides a reasonable explanation of the overall ELM dynamics.

We need to discuss the effect of $\gamma_{N}.$ 
Our expectation is that the stability of the zero solution is crucial to determine the long-time dynamics
of $P(t,x)$ for a fixed $A \gg 1$. In consideration of the analysis result in Ref. [\onlinecite{2017arXiv170608036O}], it is natural to think that $P(t,x)$ will oscillate nonlinearly if the zero solution is unstable, but converge to $0$ if the zero solution is stable. 
For this prediction, we linearized Eq.~(\ref{main}) around the zero solution $P=0$ and proved that the stability of the zero solution is independent of $\gamma_{N}$, as expected:%
\begin{equation}
\partial_{t}P_{L}=iAW_{K}\left(  x\right)  P_{L}+\gamma_{L}P_{L}+\eta\partial^2_x
P_{L}. \label{lp}%
\end{equation}
Accordingly, it is reasonable to expect that $\gamma_{N}$ cannot affect the
long-time behavior of the zero solution for large $A>0$. Conversely, $\gamma_{N}$ is expected to affect the long-time behavior of the non-zero solution for large $A>0$. Under this prediction, we confirmed numerically that $\gamma_{N}$ does not affect the qualitative long-time behavior of the solutions illustrated in Fig. \ref{det} $(a)$-$(b)$. Instead, $\gamma_{N}$ can affect the amplitude of nonlinear oscillations. 
The change of the amplitude $\left\vert P\left(  t,0\right)
\right\vert $ in our model is strongly associated with $\left(  \gamma
_{L}/\gamma_{N}\right)  ^{1/2}=\frac{1}{a}\left(  \frac{p_{\text{ref}}}%
{\gamma_{L0}}\left(  Q-Q_{c}\right)  \right)  ^{1/2}.$

\subsection{Long-time behavior of $P\left(  t,x\right)  $ on $A$ and $K$}%

Fig.~\ref{peak} suggests that there exists a threshold flow shear amplitude $A_{K}$ 
for given $K$ for both boundary conditions. 
If $0<A<A_{K}$ (blue regions), the solution $P(t,x)$ converges to a nonconstant steady state
$P_{s}\left(  x\right)  $ for any given initial condition. 
On the other hand, the qualitative long-time behavior of $P(t,x)$ abruptly changes if $A>A_{K}$ (red regions). $P(t,x)$ oscillates nonlinearly and never converges to any steady state in the red regions. 
These numerical results show that there is a certain stability/instability criterion $A_{K}$ of $A$ for
each $K>0$ for both boundary conditions. 
According to Fig.~\ref{peak}, we can also predict that ELM crash only occurs under sufficiently strong flow shear. We can also observe that as approaching the threshold line in Fig.~\ref{peak}, the amplitude of nonlinear oscillations (in the red regions) increases and the central value $\vert P(0) \vert$ for a nonzero steady state $P(x)$ (in the blue regions) decreases but remain finite (i.e. nonzero).
Besides, it is also observed that $A_{K}$ and $K$ are inversely correlated for small $K$ for
both boundary conditions, but $A_{K}$ barely changes for large $K$.

Mathematical clues for the two different dynamic behaviors illustrated in Figs.~\ref{det} and \ref{peak} can be explained in the case of the Neumann boundary condition. 
Let $P(t,x) = R(t,x) \exp\left(  i\theta(t,x)  \right)$ to rewrite
Eq.~(\ref{main}) as:%
\begin{align}
\partial_{t}R  &  =\gamma_{L}R+\eta\partial_{x}^{2}R-\eta R\theta'^{2}%
-\gamma_{N}R^{3},\label{R}\\
\partial_{t}\theta &  =\eta\partial_{x}\theta'+2\eta(\partial_{x}\ln
R)\theta'-AW_{K}\left(  x\right)  , \label{T}%
\end{align}
where $\theta'= \partial_x\theta$.
In Eq.~(\ref{R}), the shear term $AW_{K}\left(  x\right)$ affects the amplitude $R$ only indirectly via the phase-gradient $\theta'$. 
Without flow-shear ($A=0$), the steady-state $P=\left(  \gamma_{L}/\gamma_{N}\right)^{1/2}$ is the only stable equilibrium [\onlinecite{jimbo1994stability}].
Hence, without flow-shear, the phase-gradient $\theta'$ converges to $0$. 
However, for finite flow-shear, the term  $\eta R \theta'^{2}$ in Eq.~(\ref{R}) is nonzero and causes $R$ to decay in time.
If the shear is large, the term $\eta R\theta'^{2}$ dominates the linear growth term
$\gamma_{L}R$ in a neighborhood of $x=0$, so $R\left(  t,0\right)  $ decays due to the phase-gradient $\theta'$ until a critical phase-gradient $\theta'=\theta'_c$ is reached. 
After decaying, however, the term $\gamma_{N}R^{3}$ is weak close to $0$ 
and the term $\eta\partial_{x}^{2}R$ grows so large that $R(t,0)$ tends to return to its original state with the help of the linear drive $\gamma_{L}R$. 
This interaction between decay and growth terms makes the nonlinear oscillation.
However, if $\gamma_{L}$ is too small, i.e., the mode is linearly stable, the term
$\eta\partial_{x}^{2}R$ is insufficient to fully dominate the term $\eta R\theta'^{2}$. 
Accordingly, it is impossible to return to the initial state and $R(t,x)$ converges to $0$ instead. 
Similar explanations for the behavior of nonlinear oscillations were introduced in [\onlinecite{leconte2016ginzburg},\onlinecite{2017arXiv170608036O}].
In addition, it can be proved that $K$ is not an important parameter in Fig.~\ref{peak} for $K\gg 1$ [c.f. Appendix]. 

\subsection{The effect of time-varying $A$}

\begin{figure}
[ptb]
\begin{center}
\includegraphics[
width=\textwidth,
keepaspectratio
]%
{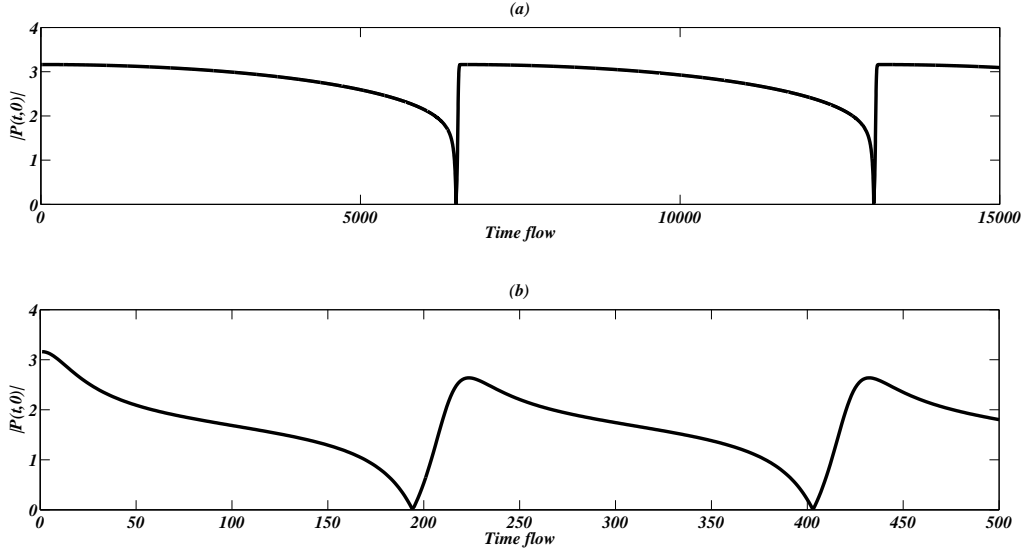}%
\caption{The time behaviors of the amplitude $\left\vert P(t,0)\right\vert$ of the solution $P(t,x)$ to Eq.~(\ref{main}) with
$\gamma_{N}=1,$ $\gamma_{L}=10,$ $\eta=1$, $W\left(  x\right)  =\tanh(25x)$ on
the Neumann boundary condition.
The initial condition is $P(0,x)=\left(  \gamma_{L}/\gamma
_{N}\right)  ^{1/2}\cos\left(  \frac{\pi x}{2}\right) $. 
(a) $A(t)$ is modeled such that $A$ increases linearly on time from $0$ initially but decreases to $0$ rapidly after the transition (crash) which occurs at $A\approx 6.5$, and this procedure is repeated. 
(b) $A = 6.5$ is constant. The quasi-steady state is only observed in (a).}%
\label{sr1}%
\end{center}
\end{figure}
\begin{figure}
[ptb]
\begin{center}
\includegraphics[
width=\textwidth,
keepaspectratio
]%
{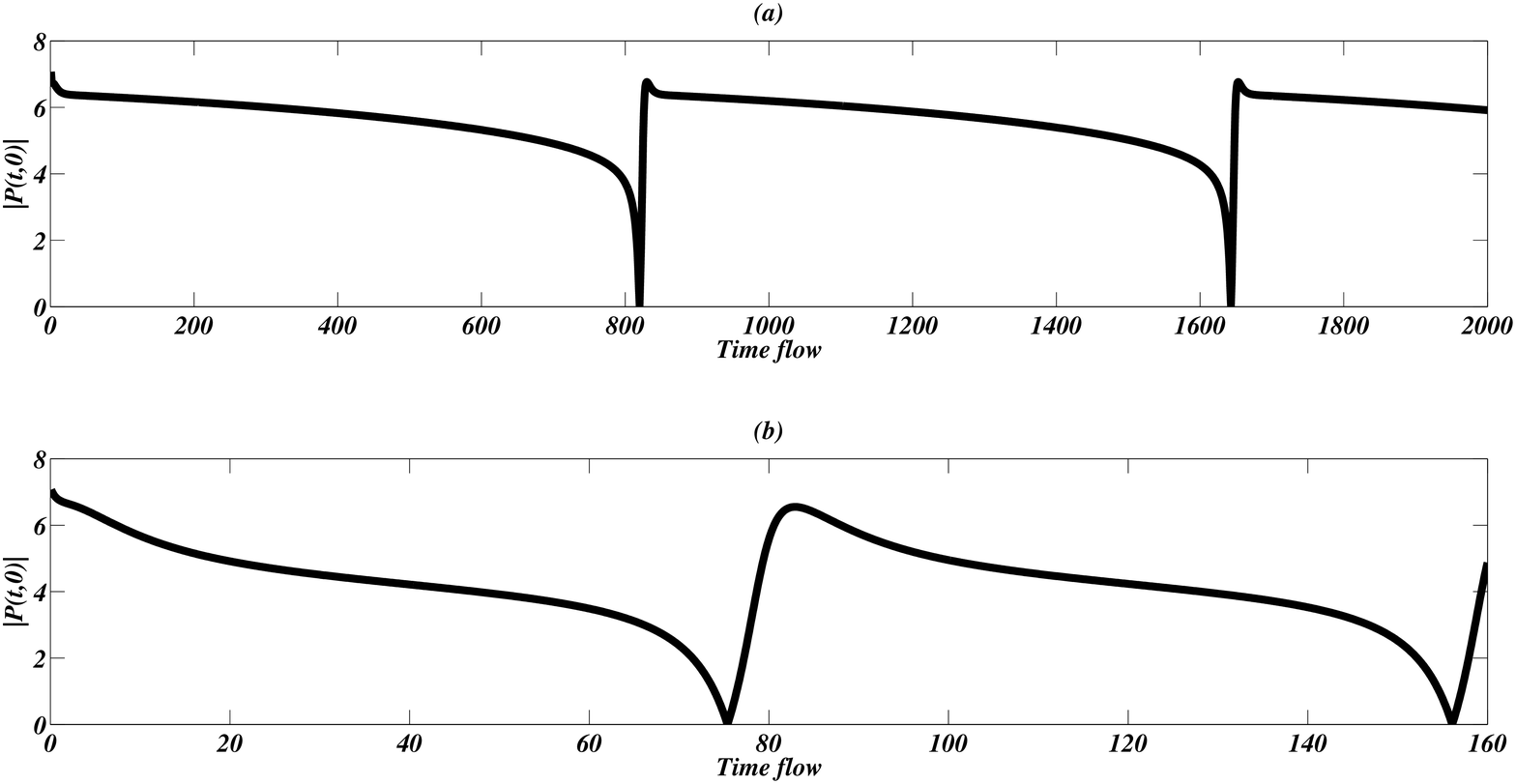}%
\caption{The time behaviors of the amplitude $\left\vert P(t,0)\right\vert$ of the solution $P(t,x)$ to Eq.~(\ref{main}) with
$\gamma_{N}=1,$ $\gamma_{L}=50,$ $\eta=1$, $W\left(  x\right)  =\tanh(25x)$ on
the Dirichlet boundary conditions. 
The initial condition is $P(0,x)=\left(  \gamma_{L}/\gamma
_{N}\right)  ^{1/2}\cos\left(  \frac{\pi x}{2}\right)$. 
(a) $A(t)$ is modeled such that $A$ increases linearly on time from $10$ initially but decreases
to $10$ rapidly after the transition (crash) which occur at $A\approx 18$, and this procedure is repeated. 
(b) $A=18$ is constant. The quasi-steady state is only observed in (a).}%
\label{sr2}%
\end{center}
\end{figure}

Nevertheless, we could not observe non-oscillating quasi-steady state for
the prescribed shear flow $AW_{K}(x)$ for both boundary conditions when $A>A_K$. 
The existence of a quasi-steady state is important for the validation of our model because the ELM dynamics observed on the KSTAR consists of distinctive stages including quasi-steady states, transition phase, and crash phase [\onlinecite{yun2011two}]. 
We believe that it is impossible to obtain a quasi-steady state for time-independent coefficients. 
Indeed, if $\left\vert \partial P_{L}/\partial t\right\vert \ll 1$, then a solution should be close to a steady
state. However, there is no reasonable steady state $P_{A,K}^{s}\ $(such that
$\partial_{x}P_{A,K}^{s}\left(  x\right)  \leq0$ in $0 \leq x\leq 1$ and
$\partial_{x}P_{A,K}^{s}\left(  x\right)  \geq0$ in $-1 \leq x\leq 0$) for a
sufficiently large fixed shear [\onlinecite{2017arXiv170608036O}], so we cannot expect
a quasi-steady state. In real experiment, it is natural to think that shear
flow evolves, i.e., $A$ and $K$ vary in time. Thus, it makes
sense that in the quasi-steady state phase, the parameters $A$ and $K$ are
initially located in a region where solutions converge to a steady state (blue
regions in Fig.~\ref{peak}), but as time flows, a shear flow
gradually increases, and $A$ and $K$ gradually change. As a consequence, as $A$
exceeds the critical point $A_{K}$, i.e., $A$ moves from the blue regions to
the red regions in Fig.~\ref{peak}, the quasi-steady state can no longer exist, which may amount to the sudden crash observed in each ELM cycle.%

The existence of quasi-steady states with time-varying $A(t)$ is numerically illustrated in Figures \ref{sr1} and \ref{sr2} for both boundary conditions. These numerical examples suggest that the change
of $A$ induces different stages in the ELM dynamics. 
Based on these results, we expect that magnetic perturbations can reduce the shear flow strength $A$ such that quasi-steady ELMs can persist without crash, which would correspond to the suppression (absence) of ELM crashes.

\section{Analysis of coupled modes}

In this section, we consider two coupled modes with the Neumann boundary condition to study the mode transitions during the quasi-steady observed on the KSTAR~[\onlinecite{lee2015toroidal}]. 
Let $W\left(  x\right)  $ be a prescribed shear flow profile and the pressure $P$ be written
as
\[
P=\overline{P}+\widetilde{P},
\]
where $\overline{P}=\overline{P}(t,x)$ is the slowly time-varying equilibrium pressure and $\widetilde
{P}=\widetilde{P}(t,x,y)$ is the pressure perturbation:%
\begin{equation}
\widetilde{P}=P_{1}\exp\left(  ik_{1}y \right)  + P_{2}\exp\left(  ik_{2}y\right) +c.c.
, \label{0}%
\end{equation}
with $\left\vert k_{1}\right\vert \neq\left\vert k_{2}\right\vert $. Extending the single mode model in
[\onlinecite{leconte2016ginzburg}], we consider the following model:%
\begin{align}
\frac{\partial P_{1}}{\partial t}-\eta\frac{\partial^{2}P_{1}}{\partial x^{2}%
}+i k_1 AW\left(  x\right) P_{1}  &  =-b\left(
\frac{\partial\overline{P}}{\partial x}P_{1}\right)  +C_{1}P_{1},\label{1-1}\\
\frac{\partial P_{2}}{\partial t}-\eta\frac{\partial^{2}P_{2}}{\partial x^{2}%
}+i k_2 AW\left(  x\right) P_{2} &  =-b\left(
\frac{\partial\overline{P}}{\partial x}P_{2}\right)  +C_{2}P_{2},\label{1-2}\\
\frac{\partial\overline{P}}{\partial t}+c\frac{\partial}{\partial x}\left(
\int_{0}^{1} \vert \widetilde{P} \vert ^{2} dy\right)   &
=d\frac{\partial^{2}\overline{P}}{\partial x^{2}}, \label{1-3}%
\end{align}
where $\eta>0,$ $A>0,$ $b>0$, $c>0$, $d>0,$ $C_{1}\geq0,$ and $C_{2}\geq0$ are
constants. With the help of the slaving approximation $\left(  \frac
{\partial\overline{P}}{\partial t}\approx0\right)$[\onlinecite{leconte2016ginzburg}],
we can obtain
\begin{equation}
\frac{c\left(  \left\vert P_{1}\right\vert ^{2}+\left\vert P_{2}\right\vert
^{2}\right)  -e}{d}=\frac{\partial\overline{P}}{\partial x}\label{1-4}%
\end{equation}
from Eq.~(\ref{1-3}) for a constant $e\in\mathbb{R}$ using 
$
\int_{0}^{1} \vert \widetilde{P} \vert^{2}dy=\left\vert P_{1}\right\vert
^{2}+\left\vert P_{2}\right\vert ^{2}.
$
Therefore, substituting Eq.~(\ref{1-4}) into Eqs.(\ref{1-1})-(\ref{1-2}) yields%
\begin{align}
\frac{\partial P_{1}}{\partial t}-\eta\frac{\partial^{2}P_{1}}{\partial x^{2}}%
+iAk_{1}W\left(  x\right)  P_{1}  &  =-b\left(  \frac{c\left(  \left\vert
P_{1}\right\vert ^{2}+\left\vert P_{2}\right\vert ^{2}\right)  -e}{d}\right)
P_{1}+C_{1}P_{1},\label{1-5}\\
\frac{\partial P_{2}}{\partial t}-\eta\frac{\partial^{2}P_{2}}{\partial x^{2}}%
+iAk_{2}W\left(  x\right)  P_{2}  &  =-b\left(  \frac{c\left(  \left\vert
P_{1}\right\vert ^{2}+\left\vert P_{2}\right\vert ^{2}\right)  -e}{d}\right)
P_{2}+C_{2}P_{2}, \label{1-6}%
\end{align}
Denoting%
\begin{align*}
\gamma_{N}  &  :=\frac{bc}{d},\\
\gamma_{L_{1}}  &  :=\left(  \frac{be}{d}+C_{1}\right)  ,\\
\gamma_{L_{2}}  &  :=\left(  \frac{be}{d}+C_{2}\right)  ,
\end{align*}
we can rewrite Eqs.(\ref{1-5})-(\ref{1-6}) as%
\begin{align}
\frac{\partial P_{1}}{\partial t}-\eta\frac{\partial^{2}P_{1}}{\partial x^{2}}%
+iAk_{1}W\left(  x\right)  P_{1}  &  =-\gamma_{N}P_{1}\left(  \left\vert P_{1}\right\vert
^{2}+\left\vert P_{2}\right\vert ^{2}\right)  +\gamma_{L_{1}}P_{1},\label{1-7}\\
\frac{\partial P_{2}}{\partial t}-\eta\frac{\partial^{2}P_{2}}{\partial x^{2}}%
+iAk_{2}W\left(  x\right)  P_{2}  &  =-\gamma_{N}P_{2}\left(  \left\vert P_{1}\right\vert
^{2}+\left\vert P_{2}\right\vert ^{2}\right)  +\gamma_{L_{2}}P_{2}. \label{1-8}%
\end{align}
Let $P_{1}=R_{1}\exp\left(  i\theta_{1}\right)  $ and $P_{1}=R_{2}\exp\left(
i\theta_{2}\right)  .$ Then Eqs.(\ref{1-7})-(\ref{1-8}) can be written as%
\begin{align*}
\dot{R}_{1}-\eta R_{1}^{\prime\prime}+\eta R_{1}{\theta_1^\prime}^2  &
=-\gamma_{N}\left(  R_{1}^{3}+R_{1}R_{2}^{2}\right)  +\gamma_{L_{1}}R_{1},\\
\dot{R}_{2}-\eta R_{2}^{\prime\prime}+\eta R_{2}{\theta_2^\prime}^2  &
=-\gamma_{N}\left(  R_{2}^{3}+R_{1}R_{2}^{2}\right)  +\gamma_{L_{2}}R_{2}%
\end{align*}
We assume that $\gamma_{L_{1}}\neq\gamma_{L_{2}}.$ Here, we can interpret
$\gamma_{N}$, $\gamma_{L_{1}},$ $\gamma_{L_{2}}$ and $\eta$ as constant
coefficients for the nonlinear term, the linear growth terms for $P_{1}$ and $P_{2}$, and the dissipative term
respectively. In this paper, we only consider positive values of $\gamma_{L_{1}%
},$ $\gamma_{L_{2}},$ $\eta$ and $\gamma_{N}.$ The only difference from
Eq.~(\ref{main}) to Eqs.~(\ref{1-7})-(\ref{1-8}) is the presence of the coupling terms $\gamma
_{N}P_{1}\left\vert P_{2}\right\vert ^{2}$ and $\gamma_{N}P_{2}\left\vert P_{1}\right\vert
^{2}$ in the equations for $P_{1}$ and $P_{2}$ respectively, which can account for the mode transition observed in [\onlinecite{lee2015toroidal}].

\subsection{Long-time behavior on the linear growth terms}
\begin{figure}
[ptb]
\begin{center}
\includegraphics[
width=\textwidth,
keepaspectratio
]%
{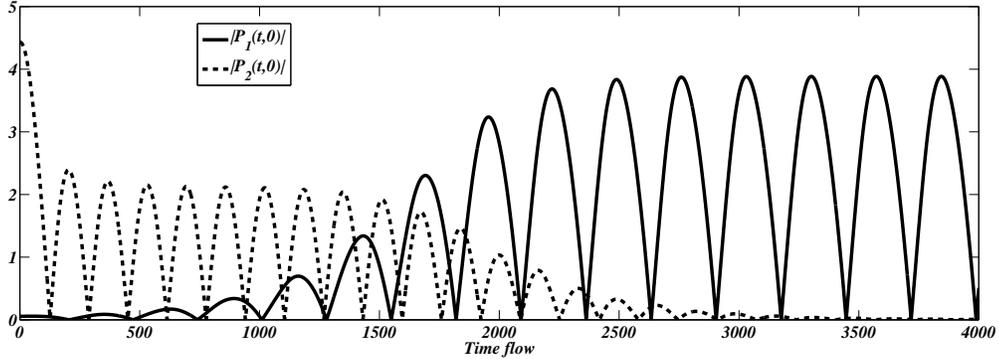}%
\caption{The time behaviors of $\left\vert P_{1}(t,0)\right\vert $ and $\left\vert
P_{2}(t,0)\right\vert$ where $P_{1}(t,x)$ and $P_{2}(t,x)$ are solutions to Eqs.(\ref{1-7}),(\ref{1-8})
respectively. We set $\eta=1$, $A=10,$ $k_{1}=5,$
$k_{2}=8,$ $\gamma_{N}=1$ and $W\left(  x\right)  =\tanh(25x).$ Besides, we
imposed $\gamma_{L_{1}}=30$ and $\gamma_{L_{2}}=20$ and initial conditions $P_{1}(0,x)$ and $P_{2}(0,x)$
 as $\left( \frac{\gamma_{L_1}}{\gamma_{N}}\right)  ^{1/2}\left(  0.01\right)  $ and $\left(
\frac{\gamma_{L_2}}{\gamma_{N}}\right)  ^{1/2}\left(  0.99\right)  $ respectively. 
$\left\vert P_{1}(t,0)\right\vert $ becomes dominant and oscillates nonlinearly although the initial value is small while $\left\vert P_{2}(t,0)\right\vert $ converges to
$0$ although the initial value is large. Hence, the conditions
$\gamma_{L_{1}}>\gamma_{L_{2}}$ and $k_{1}<k_{2}$ means the dominance of
$\left\vert P_{1}(t,0)\right\vert$ for sufficiently large shear.}%
\label{mt01}%
\end{center}
\end{figure}

\begin{figure}
\begin{center}
\includegraphics[
width=\textwidth,
keepaspectratio
]%
{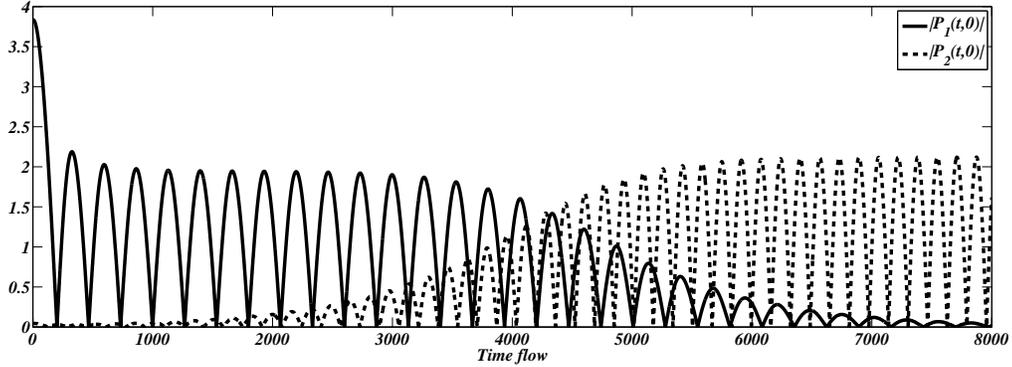}%
\caption{The time behaviors of $\left\vert P_{1}(t,0)\right\vert $ and $\left\vert
P_{2}(t,0)\right\vert $ where $P_{1}(t,x)$ and $P_{2}(t,x)$ are solutions to Eqs.(\ref{1-7}),(\ref{1-8})
respectively. We set the same values for the parameters $\eta, A, k_{1},k_{2}, \gamma_{N}$, and $W(x)$ as in Fig.~\ref{mt01}. Besides, we imposed $\gamma_{L_{1}}=15$ and $\gamma_{L_{2}}=20$ and initial conditions $P_{1}(0,x)$ and $P_{2}(0,x)$ as $\left(  \frac{\gamma_{L_1}}{\gamma_{N}}\right)  ^{1/2}\left(  0.99\right)  $ and $\left(
\frac{\gamma_{L_2}}{\gamma_{N}}\right)  ^{1/2}\left(  0.01\right)  $ respectively. 
$\left\vert P_{1}(t,0)\right\vert $ converges to $0$
and $\left\vert P_{2}(t,0)\right\vert $ oscillates nonlinearly, showing that the linear growth terms highly affect the long-time behavior of the two modes.}%
\label{mt02}%
\end{center}
\end{figure}

To understand the dependence of the time behavior of the couple modes on the linear growth terms, 
we performed numerical calculations with 
fixed $\eta=\gamma_{N}=1, A=10, W(x)=\tanh(25x)$, $k_{1}=5$ and $k_{2}=8$ for different $\gamma_L's$. 
Fig.~\ref{mt01} shows the time behaviors of
$\left\vert P_{1}(t,0)  \right\vert $ and $\left\vert P_{2}(t,0)  \right\vert $ for $\gamma_{L_{1}}=30$ and $\gamma_{L_{2}}=20$  with the initial condition $\left\vert P_{1}(0,x) \right\vert \ll \left\vert P_{2}(0,x) \right\vert$.
$\vert P_{1}(t,0) \vert$ grows and becomes dominant with nonlinear oscillation while $\vert P_{2}\left(  t,0\right)\vert$ decays. 
Fig.~\ref{mt02} shows the case for $\gamma_{L_{1}}=15$ and $\gamma_{L_{2}}=20$ 
with the opposite initial condition $\left\vert P_{1}(0,x) \right\vert \gg \left\vert P_{2}(0,x) \right\vert$. 
$\left\vert P_{1}(t,0) \right\vert $ converges to $0$ and
$\left\vert P_{2}(t,0)  \right\vert $ becomes dominant as $t\rightarrow \infty$. 

In both cases, the mode with higher $\gamma_L$ becomes dominant eventually as expected.
However, there is a subtle difference in the time scale between Fig.~\ref{mt01} and Fig.~\ref{mt02}.
We can explain the difference as follows.
For the case of Fig.~\ref{mt01}, $\gamma_{L_{1}}>\gamma_{L_{2}}$ and
$\left\vert k_{1}\right\vert <\left\vert k_{2}\right\vert $ mean that $P_{1}$ has
stronger linear growth and, at the same time, less suppression due to the shear compared to $P_{2}$ so that $P_1$ will quickly become dominant. 
However, in the case of Fig.~\ref{mt02}, 
although $\gamma_{L_{1}}<\gamma_{L_{2}}$, 
it takes longer for $P_2$ to become dominant because $P_{1}$ is less suppressed than $P_{2}$ by the shear.

To conclude, the long-time behaviors of $\left\vert P_{1}\right\vert $ and
$\left\vert P_{2}\right\vert $ under `fixed' parameters with $k_{1}<k_{2}$ are
determined by $\gamma_{L_1}$ and $\gamma_{L_2}$.

\begin{figure}
[ptb]
\begin{center}
\includegraphics[width=\textwidth,
keepaspectratio
]%
{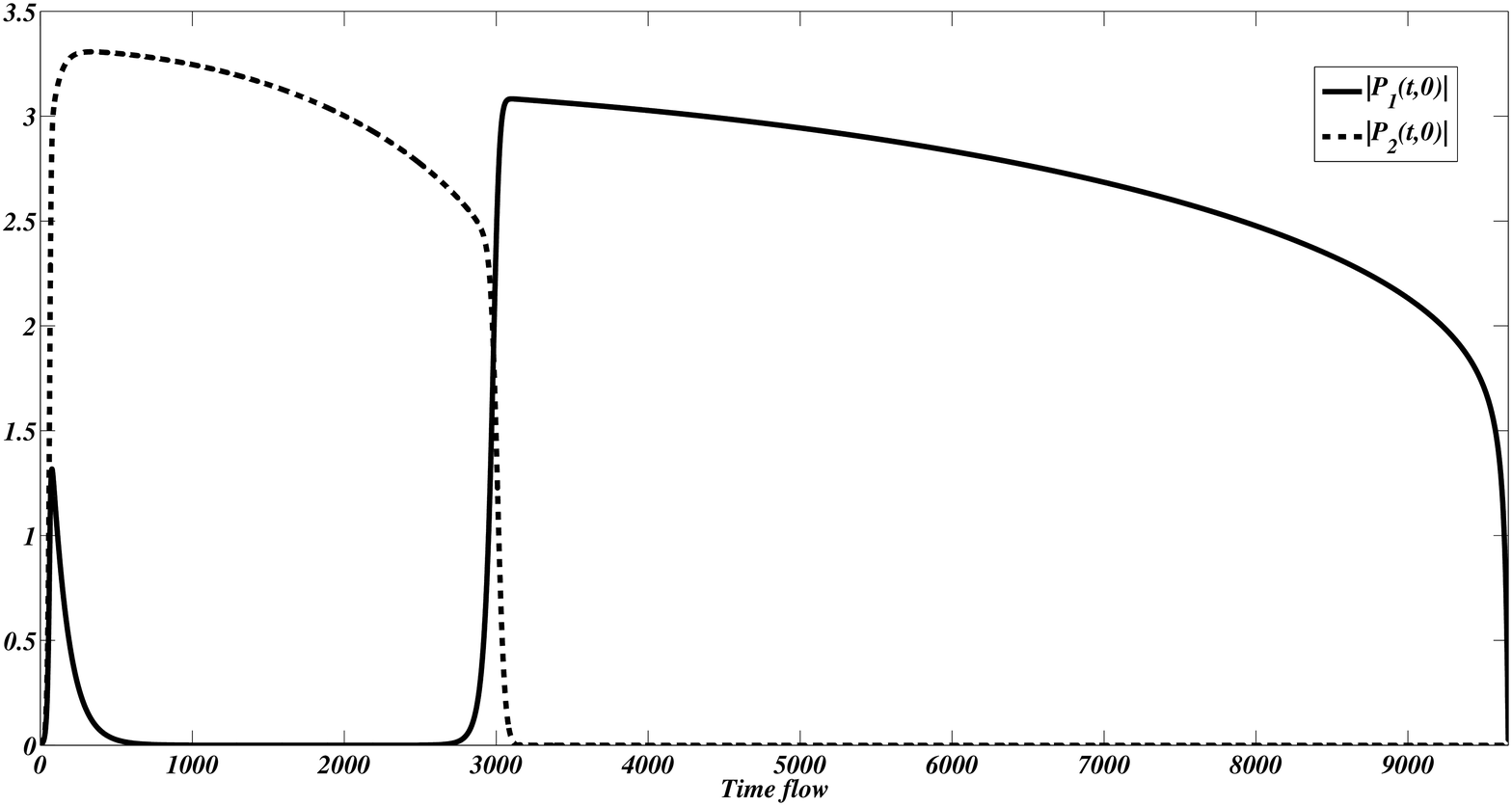}%
\caption{The time behaviors of $\left\vert P_{1}(t,0)\right\vert $ and $\left\vert
P_{2}(t,0)\right\vert $ for time-dependent $A(t)$ where $P_{1}(t,x)$ and $P_{2}(t,x)$ are solutions to Eqs.(\ref{1-7}),(\ref{1-8})
respectively with $\eta=1$, $k_{1}=1,$ $k_{2}=3,$ $\gamma_{L_{1}}=10$, $\gamma_{L_{2}}=11$, $\gamma_{N}=1$ and $W\left(  x\right)
=\tanh(25x).$ We imposed weak initial conditions $P_{1}(0,x)$ and $P_{2}(0,x)$ as $\left(  \gamma_{L_1}/\gamma_{N}\right)  ^{1/2}/1000$
and $\left(  \gamma_{L_2}/\gamma_{N}\right)  ^{1/2}/1000$ respectively. $A$ increases linearly, reaching the value $6.4393$ at the end of the horizontal $x$-axis in the figure. 
First, $P_{2}(t,0)$ is dominant and quasi-steady when the shear is small. As the shear increases beyond a critical value, $\left\vert P_{1}(t,0) \right\vert $ increases rapidly while $\left\vert P_{2}(t,0) \right\vert$ vanishes rapidly. After that, $P_{1}(t,0)$ remains in a quasi-steady state until it falls to $0$ abruptly.}
\label{sr3}%
\end{center}
\end{figure}

\begin{figure}
[ptb]
\begin{center}
\includegraphics[width=\textwidth,
keepaspectratio
]%
{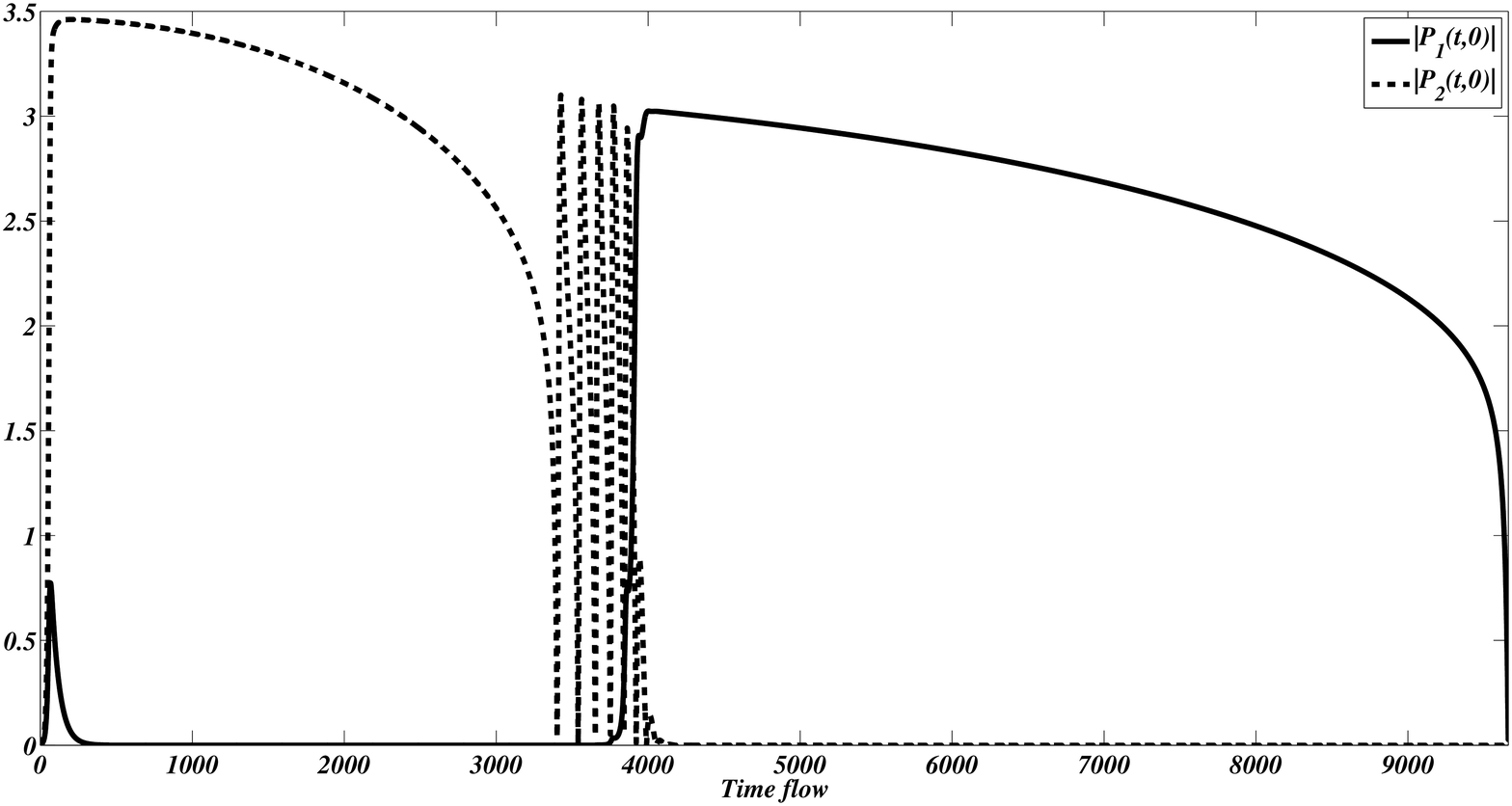}%
\caption{The time behaviors of $\left\vert P_{1}(t,0)\right\vert $ and $\left\vert
P_{2}(t,0)\right\vert $ for time-dependent $A(t)$ where $P_{1}(t,x)$ and $P_{2}(t,x)$ are solutions to Eqs.(\ref{1-7}),(\ref{1-8})
respectively with $\eta=1$, $k_{1}=1,$ $k_{2}=3,$ $\gamma_{L_{1}}=10$, $\gamma_{L_{2}}=12$, $\gamma_{N}=1$ and $W\left(  x\right)
=\tanh(25x).$ We imposed weak initial conditions $P_{1}(0,x)$ and $P_{2}(0,x)$ as $\left(  \gamma_{L_1}/\gamma_{N}\right)  ^{1/2}/1000$
and $\left(  \gamma_{L_2}/\gamma_{N}\right)  ^{1/2}/1000$ respectively. $A$ increases linearly, so $A$ reaches to $6.4393$ at the end of the horizontal $x$-axis in the figure. First, $\left\vert
P_{2}(t,0)\right\vert $ is dominant when the shear is small. As the shear increases,
$\left\vert P_{1}(t,0)\right\vert $ increases, but $\left\vert P_{2}(t,0)\right\vert $
decreases. After that, $\left\vert P_{1}(t,0)\right\vert $ act as a quasi-steady state,
and finally, $\left\vert P_{1}(t,0)\right\vert $ falls to $0$ abruptly. Compared to Fig. \ref{sr3}, it is also remarkable that the high oscillation of $\left\vert P_{2}(t,0)\right\vert $ before
converging to $0$ is observed.}%
\label{sr4}%
\end{center}
\end{figure}

\subsection{Long-time behavior for time-varying $A.$}

The analysis shown in Figs.
\ref{mt01}-\ref{mt02} still cannot explain the transitions between quasi-stable modes observed in experiments [\onlinecite{lee2015toroidal}]. Now, we consider time-varying $A$ in Eqs.(\ref{1-7})-(\ref{1-8}) to understand the mode transitions for the case with  $\gamma_{2}>\gamma_{1}$ and  $k_{2}>k_{1}$.
$P_{2}$ is dominant for sufficiently small $A$. 
If $A$ increases in time, it is expected that $P_{2}$ is more suppressed than $P_{1}$ because $k_{2}>k_{1}$ means
that $P_{2}$ is more sensitive to $A$ than $P_{1}$, so $P_{1}$ can become dominant finally.
Figs. \ref{sr3}-\ref{sr4} show the behaviors of $\left\vert P_{2}(t,0)\right\vert$ and $\left\vert
P_{1}(t,0)\right\vert$ with growing $A$, supporting our prediction. Note that $\left\vert
P_{2}(t,0)\right\vert $ is highly oscillating before convergence to $0$ in Fig.~\ref{sr4}, but not in Fig.~\ref{sr3}. We should mention that the numerical examples presented here capture the importance of time-varying $A$ and offer qualitative explanations for various types of mode transitions observed in experiments.

\section{\label{sec3}Conclusion}

In summary, we considered two cases of ELM dynamics based on the generalized Ginzburg-Landau model, Eq.~(\ref{main}). 
In the case of the single-mode, we studied the long-time behavior of the solution with fixed model coefficients and showed that $\gamma_L$ and $A$ determine the long time behavior of the solution. 
If the linear growth term is sufficiently large, the
nonlinear oscillations are guaranteed for large shear flow. Conversely, the solution converges to a nonzero steady state for weak shear flow (Fig.~\ref{peak}).
The long-time behavior for the small linear growth
term is interesting because a solution converges to $0$ for large flow shear (Fig.~\ref{det}). 
Combining these results, we conclude that it is insufficient to consider the fixed coefficients on time
to realize the quasi-steady states which are observed in experiments [\onlinecite{yun2011two}].
Therefore, by imposing time-varying shear flow, we obtained quasi-steady states numerically (Figs. \ref{sr1}-\ref{sr2}).

To study the dynamics of coupled modes $P_{1}$ and $P_{2}$, we derived equations (\ref{1-7})-(\ref{1-8}). 
We confirmed that the linear growth terms are crucial
to determine the long-time behavior of $P_{1}$ and $P_{2}$ (Figs.\ref{mt01}-\ref{mt02}). 
Inspired by these results, we considered the increasing $A(t)$ on time and showed that rapid mode transition occurs (Figs.\ref{sr3}-\ref{sr4}), reproducing qualitatively the observed mode transitions in experiments [\onlinecite{lee2015toroidal}].

Although we dealt with the equations (\ref{1-7})-(\ref{1-8}) for coupled-modes, it is also possible to obtain equations for more than two modes and show that each mode solution is successively dominant with suitable time-dependent $A$.

To conclude, it is critical to consider the time-varying $A$ for explanation of dynamic features in ELM phenomena using the given models (\ref{main}) and (\ref{1-7})-(\ref{1-8}) for single and coupled-modes, respectively.
Based on our numerical analysis, we expect that the quasi-stable mode can persist if the flow-shear is reduced below the critical threshold by application of external magnetic perturbations, which may provide a candidate mechanism for the non-bursting quasi-stable modes in the ELM crash suppression experiment[\onlinecite{mp3}].

\section*{Acknowledgement}
Hyung Ju Hwang was partly supported by the Basic Science Research Program
through the National Research Foundation of Korea (NRF) (2015R1A2A2A0100251).
M. Leconte was supported by R\&D Program through National Fusion Research Institute (NFRI) funded by
the Ministry of Science, ICT and Future Planning of the Republic of Korea (NFRI-EN1741-3).
Gunsu S. Yun was partially supported by the National Research Foundation of
Korea under grant No. NRF-2017M1A7A1A03064231 and by Asia-Pacific Center for
Theoretical Physics.

\section*{Appendix: Explanation of why the nonlinear oscillation threshold is independent of $K$, for large $K$}
Notice that even if
the shear $AW_{K}\left(  x\right)  $ appears, there exists a unique linearly
stable steady state denoted by the superscript $s$, $P_{A,K}^{s}=R_{A,K}^{s}\exp\left(  i\theta_{A,K}%
^{s}\right)  $ such that $R_{A,K}^{s}\left(  -x\right)  =R_{A,K}^{s}\left(
x\right)  $ and $\partial_{x}\theta_{A,K}^{s}\left(  x\right)  =\partial
_{x}\theta_{A,K}^{s}\left(  -x\right)  $ for small $A<<1$
 [\onlinecite{2017arXiv170608036O}]$.$ We can also deduce from (\ref{T})
\begin{equation}
\frac{\partial \theta}{\partial x} 
\Big|_{A,K}^{s}  =\frac{A}{\eta}\int_{-1}^{x}W_{K}\left( x' \right)
\frac{R_{A,K}^{s}\left( x' \right)  }{R_{A,K}^{s}\left(  x\right)
}dx'. \label{TT}%
\end{equation}
It should be checked how $K$ affects the profile of $\left\vert P_{A,K}^{s}\right\vert .$ It was numerically observed that there are stable symmetric stable
steady states before $A<A_{K}$ (see [\onlinecite{2017arXiv170608036O}]). Due
to
\[
\lim_{K\rightarrow\infty}W_{K}\left(  x\right)  =\left\{
\begin{array}
[c]{c}%
-1\text{ if }x<0\\
1\text{ if }x>0
\end{array}
\right\}  ,
\]
we can obtain%
\begin{align}
\frac{\partial \theta}{\partial x} \Big|_{A,K}^{s} &  =-A\int_{-1}^{x}\frac{R_{A,K}^{s}\left(
x'\right)  }{R_{A,K}^{s}\left(  x\right)  }dx'+A\int_{-1}^{x}\left(
W_{K}\left( x' \right)  +1\right)  \frac{R_{A,K}^{s}\left( x' \right)  }%
{R_{A,K}^{s}\left(  x\right)  }dx'\label{sakt}\\
&  \approx-A\int_{-1}^{x}\frac{R_{A,K}^{s}\left( x' \right)  }{R_{A,K}%
^{s}\left(  x\right)  }dx',\nonumber
\end{align}
if $K>>1.$ Therefore, the equation (\ref{R}) for $P_{A,K}^{s}=R_{A,K}^{s}\exp\left(
i\theta_{A,K}^{s}\right)  $ barely changes for $K>>1$, so the profile of
$\left\vert R_{A,K}^{s}\left(  x\right)  \right\vert $ is almost independent
of $K$ for $K>>1$ due to (\ref{sakt})$.$

\bibliography{manuscript}

\end{document}